# LANGMUIR-BLODGETT FILMS FROM ORGANIC SOLVENTS

*Alexander Sorokin [a] [*], Larissa Maiorova [b,c], Maksim Zavalishin [b]*

[a] Independent researcher, Ivanovo, Russia
[b] Ivanovo State University of Chemistry and Technology, Ivanovo 153000, Russia
[c] Federal Research Center "Computer Science and Control" of the Russian Academy of Sciences, Moscow 119333, Russia

\* Corresponding author.
E-mail addresses: alex@sorokin.info (Alexander Sorokin), maiorova.larissa@gmail.com (Larissa Maiorova), zavalishin00@gmail.com (Maksim Zavalishin).

*Surface layers of water miscible volatile organic solvents (N,N-dimethylacetamide, acetone, ethanol, and tetrahydrofuran) have been formed on water and their properties have been studied. The layers are stable on the water surface and can be transferred to a solid substrate to form a multilayer film. Despite the volatility of the source material, some films can even withstand heating.*

## 1. Introduction

Our recent work [1] reports the ability of water-miscible dimethyl sulfoxide (DMSO) to form stable surface layer on water. Such a layer, prepared by spreading DMSO with a special device, can be compressed by the movable barriers of a Langmuir trough and maintained under constant surface pressure for a sufficiently long time to allow multiple transfers of the layer to solid substrates by the standard Langmuir-Blodgett (LB) technique (horizontal lift method). The resulting multilayer films are stable and sparingly soluble even in hot water. It has been proposed that the artificially spread layers of DMSO are similar to the naturally adsorbed (Gibbs) layers of DMSO, but are not in equilibrium with the solution. The higher concentration of DMSO in such supersaturated layers is metastable, it persists for quite a long time because the DMSO molecule has a minimum of free energy on a surface and its transfer into the volume requires overcoming the energy barrier.

It is obvious that the proposed model of the surface layer, if true, should be valid not only for DMSO. Any water-soluble surfactant fits this model if, when spread on the water surface, it does not dissolve immediately but remains on the surface for a prolonged period of time, allowing for higher than equilibrium concentrations. Therefore, similar behavior to DMSO can be expected from other polar solvents such as acetone or tetrahydrofuran.

This assumption was tested and proven in this work. We tried N,N-dimethylacetamide, acetone, ethanol, and tetrahydrofuran and found that these common organic solvents do form surface layers on water. The multilayer films on substrates can be made even easier from these solvents than from DMSO, because the special device that facilitates spreading is not always necessary. Surprisingly, the films are quite stable even when made from volatile solvents.

## 2. Experimental

### 2.1. Solvents

Table 1 lists the solvents studied or mentioned in this paper along with their properties.

Table 1. Solvents studied or mentioned in this paper.

| Solvent | Density | Molar mass | Dipole moment | Dielectric constant | Melting point | Boiling point | Surface tension |
|---|---|---|---|---|---|---|---|
| | g/cm³ | g/mol | Debye | | °C | °C | mN/m |
| Dimethyl sulfoxide (DMSO) | 1.10 | 78 | 3.9 | 47 | +18 | 189 | 42 |
| N,N-dimethylacetamide (DMAA) | 0.94 | 87 | 3.8 | 38 | −20 | 165 | 35 |
| Tetrahydrofuran (THF) | 0.89 | 72 | 1.6 | 8 | −108 | 66 | 28 |
| Acetone | 0.79 | 58 | 2.9 | 21 | −95 | 56 | 24 |
| Ethanol | 0.79 | 46 | 1.7 | 24 | −114 | 78 | 22 |
| DMSO | DMAA | THF | Acetone | Ethanol | | | |
| 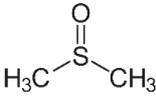 | 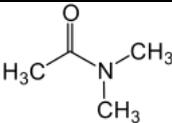 | 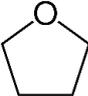 | 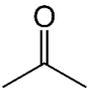 | 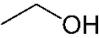 | | | |

### 2.2. The Langmuir trough and injection of solvents into it

The Langmuir setup and the water used as the subphase are described in [1].

In [1] it was reported that DMSO injected dropwise into the Langmuir trough formed the surface layer occasionally and unpredictably. This was explained by the higher density of DMSO compared to water, which causes the droplet of DMSO to enter the subphase volume rather than spread over its surface. To overcome this obstacle, the method of DMSO injection has been proposed, which involves impregnating a sheet of porous material with DMSO, removing the excess DMSO, and aligning the bottom edge of the sheet, which is suspended vertically, with the surface of the subphase. This method perfectly forms the spread surface layer of DMSO on water. Non-volatile DMSO does not evaporate from the sheet, so the size of the sheet and the time it is in contact with the surface is unlimited. Eventually, all of the DMSO that initially impregnated the sheet will end up in the Langmuir trough and its significant portion will be in the surface layer, although this may take several hours for tall sheets.

All the solvents studied in this work are volatile. They can evaporate from the upper part of the porous sheet earlier than they will be in contact with the water surface, which makes large height of the sheet impractical. For highly volatile solvents, only a narrow lower portion of the porous sheet should be impregnated, and the injection time is naturally limited by the evaporation of the solvent. This essentially deprives the porous sheet of its function of reservoir sourcing the spreading substance, while the function of slow injection of the substance in a narrow flat stream is still present. To inject a highly volatile solvent, we attached a wide and non-high (e.g., 20×4 cm²) sheet of filter paper to the apparatus shown in Fig. 1 in [1], slightly wetted the sheet with the solvent, and rapidly lowered the sheet until its lower edge touched the surface of the subphase. After all the solvent from the sheet had migrated to the trough or air, we added more solvent to the sheet from a syringe or pipette, distributing the droplets evenly across the width of the sheet. This combined injection forms long lasting surface layers of acetone, ethanol or tetrahydrofuran on water. However, for these solvents, which are lighter than water, conventional droplet injection directly onto the surface of the aqueous subphase works quite well.

All experiments with surface layers on aqueous subphase and their transfer to a substrate were performed at room temperature (17-25 °C depending on the weather, typically 19-22 °C) without temperature control. The temperature fluctuation did not exceed 2 °C per day.

### 2.3. Transfer of surface layers to a solid substrate and preparation of multilayer films

The layer formed on the surface of the aqueous subphase was repeatedly transferred to solid substrates (polished silicon, metallized glass) at constant surface pressure using the horizontal lift method. For visual comparison of the appearance of the films with different number of transfers, the multizone samples were prepared according to the procedure described in [1].

# 3. Results

## 3.1. N,N-dimethylacetamide

N,N-Dimethylacetamide (DMAA) is an industrial solvent. Of the substances studied in this work, DMAA has the closest physical properties to DMSO, and the molecules of the two compounds have a lot in common. DMAA is moderately volatile, slightly less dense than water, and miscible with water. For this work, ReagentPlus® grade (99%) DMAA was purchased from Aldrich and used as received.

As we have learned, both the porous sheet method and regular dropwise injection are suitable for forming a surface layer of DMAA on water, but the former is more effective and requires less manual labor. The methods can be combined, or DMAA can be added dropwise onto a porous sheet already in contact with the subphase. The volatility of DMAA makes high sheets impractical. For more substance in the sheet, it is better to increase the width. The dependence of the surface pressure $\pi$ on the time $t$ during the injection of DMAA from the porous sheet, shown in Fig. 1, is very similar to the corresponding dependence for DMSO (Fig. 2 in [1]). With dropwise injection, the growth $\pi$ is noticeable after the first droplets are injected. $\pi$ remains constant for a long time after the injection stops (Fig. 2). The surface pressure-area isotherm ($\pi$-$S$ isotherm) of DMAA on water is shown in Fig. 3.

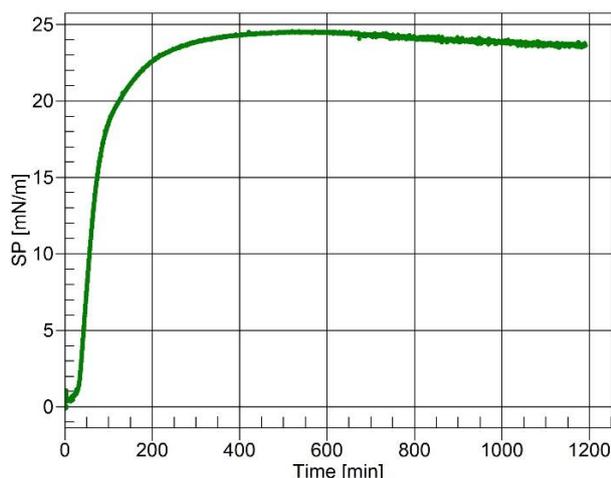

Fig. 1. Surface pressure $\pi$ versus time $t$ during and after the injection of DMAA from the porous sheet. A 20 cm (width) × 10 cm (height) sheet of filter paper was impregnated with 3.4 g of DMAA. The sheet was in contact with the surface of the aqueous subphase from $t$=0 to $t$=660 min.

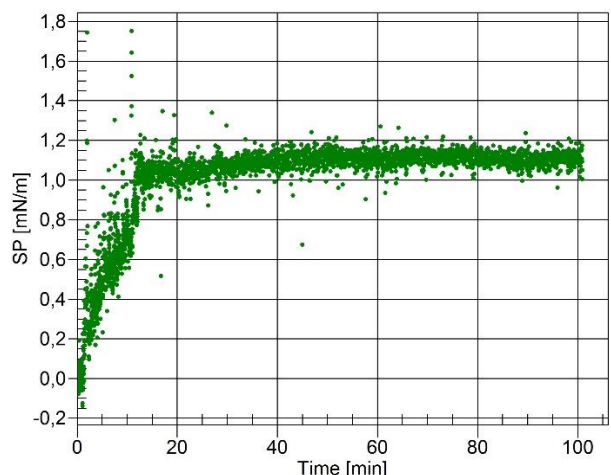

Fig. 2. Surface pressure $\pi$ versus time $t$ during and after dropwise injection of DMAA. 1 mL of DMAA was manually injected from a syringe, starting at $t$=0 and ending at $t$=12 min (droplet volume approximately 4 µL).

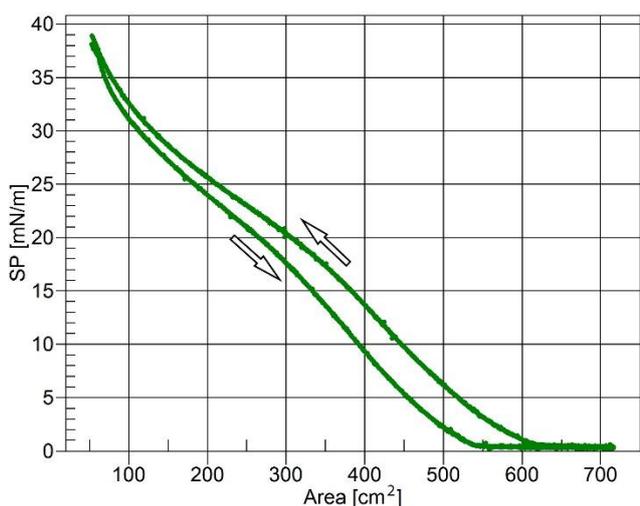

Fig. 3. The π-S isotherm of DMAA on water. The amount of DMAA was chosen experimentally so that π begins to rise very soon after the start of compression. The main amount, nearly 3 g, was injected from the porous sheet, and 0.22 g was added dropwise onto the sheet. The barriers were moved at 20 mm/min.

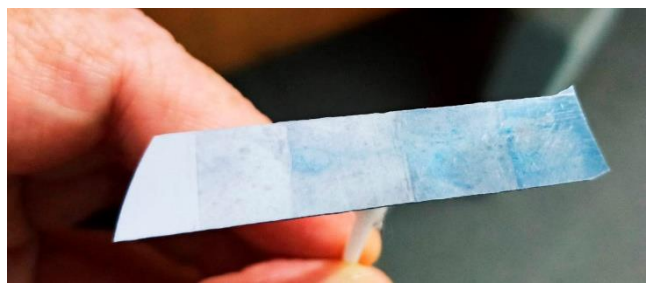

Fig. 4. The film of DMAA on the polished silicon. There are five zones on the sample with, from left to right, 0, 3, 13, 23, and 33 transferred layers. The width of the sample is 10 mm.

The surface layer of DMAA is much more stable than that of DMSO. For example, if $\pi$ is kept constant from the 15-20 mN/m range, the barriers of the Langmuir trough approach each other slower than 1 mm/hour (the area decreases by 0.5 % per hour). Similar to DMSO, the surface layer of DMAA is depleted during long-term compression ($\pi$ drops) and is restored when the pressure is released, but with DMAA the depletion is less pronounced and the restoration occurs more rapidly.

The surface layers of DMAA can be transferred to a substrate by the horizontal lift method. The transfer process is very similar to that of DMSO [1]. The surface of the film on a substrate is hydrophilic. Fig. 4 shows the multizone film on silicon, where each zone contains a different number of transferred layers.

### 3.2. Tetrahydrofuran

Tetrahydrofuran (THF) is the widely used volatile aprotic solvent, lighter than water, highly hygroscopic, miscible with water in any ratio except for a narrow concentration range at elevated temperatures [2]. THF is often used as a solvent in the preparation of Langmuir-Blodgett films, e.g. in [3–10], and the solubility of THF in water is typically not taken into account. Reference [2] indicates the amphiphilic nature of THF. When injected dropwise onto the water surface, THF does not immediately mix with the water and remains on top of it for an extended period of time. This can be observed in a test tube due to the difference in the refractive index of the liquids: the boundary between water and THF, which is sharp at first, is gradually blurred and becomes indistinguishable after several hours.

For this work, chemically pure THF was treated with copper chloride (I) to remove peroxides, distilled, and bubbled with argon.

The surface layers of THF are good with regular dropwise injections from a syringe. A small advantage of the porous sheet is faster injecting large amounts of THF; the appropriate sheet is wide and low height. The surface pressure starts to increase with the first injected droplets and changes to decrease when the injection is stopped (Fig. 5). The decrease is slow, so the isotherms start at non-zero $\pi$ (Fig. 6). If the maximum $\pi$ reached during compression is less than 25-30 mN/m, the hysteresis is weak, but at higher maximum $\pi$ the isotherm gets two inflections and a segment with a smaller slope in between.

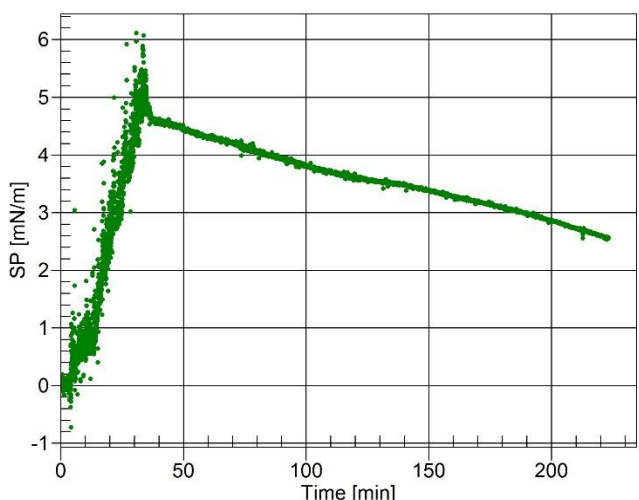
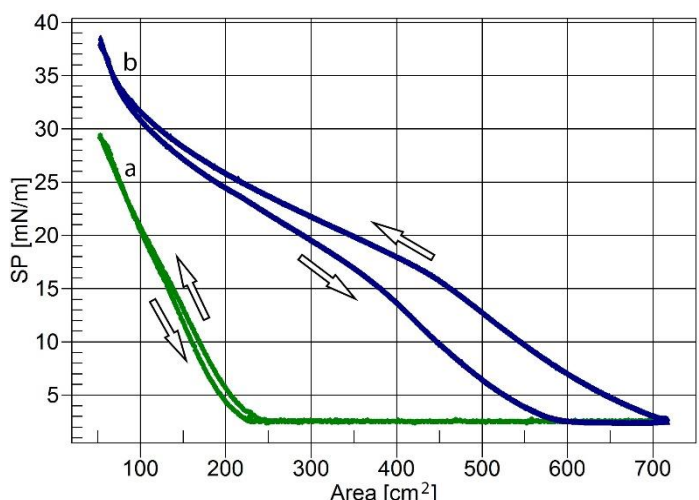

Fig. 5. Surface pressure $\pi$ versus time $t$ during and after dropwise injection of THF. 3.5 g of THF was manually injected from a glass pipette, starting at $t$=0 and ending at $t$=30 min (droplet volume approximately 30 µL).

Fig. 6. $\pi$-S isotherms of the surface layer of THF.
(a) 1.3 g of THF injected dropwise from a syringe;
(b) 2.5 g of THF injected using the porous sheet.

A freshly prepared surface layer takes time to stabilize (Fig. 5), but the layer that has been on the water for several hours shows excellent stability. For example, the surface pressure $\pi$, initially 22, 30, and 36 mN/m, decreased at the constant area by 0.15, 0.45, and 0.96%/hour (observation time at least 10 hours). When the constant $\pi$=20 mN/m was maintained, the area between the barriers decreased by only 2 % in 2 hours.

Multilayer THF films on polished silicon wafers were formed by the horizontal lift method at constant $\pi$ in the range of 9-18.5 mN/m. The substrate, when removed from the subphase surface, is quite hydrophobic and may be almost dry except for the droplet in the corner. The film adheres well to the substrate and requires some force to remove. The film immersed in water for 12 hours did not dissolve, although its appearance changed slightly, but droplets of chloroform, THF, and ethanol placed on the film destroyed it in a few minutes. Given the high volatility of the starting material, one would expect the films to be thermally unstable, but heating the sample shown in Fig. 7 at 80-110 °C for one hour did not cause any visible changes.

### 3.3. Acetone

Analytical grade (>98%) acetone was purchased from a local supplier and used as received.

Acetone behaves similarly to DMAA and THF: when injected through the surface of the subphase in a Langmuir trough, this solvent forms a long-lasting surface layer that can be compressed by barriers and repeatedly transferred to a substrate to form a multilayer film. Acetone can be injected dropwise, from a porous sheet, or a combination of the two.

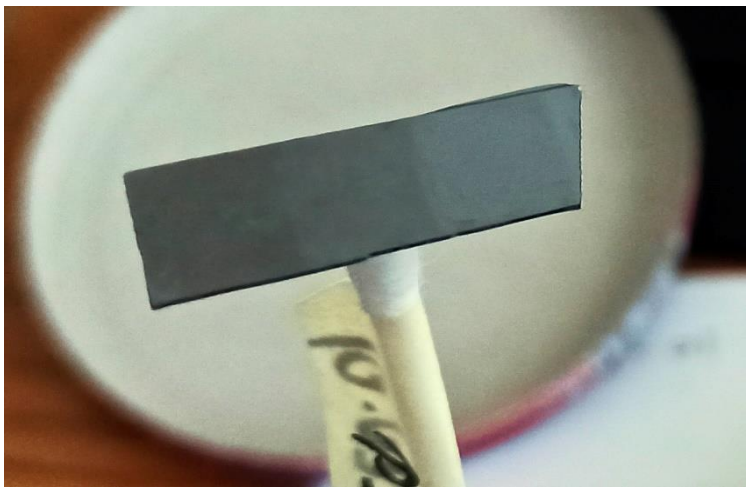

Fig. 7. The film of THF on the polished silicon. There are four zones on the sample with, from right to left, 0, 30, 60, and 90 transferred layers. The width of the sample is 10 mm.

Fig. 8 shows a series of π-S isotherms obtained after injection of about 5 g of acetone by the combined method. The first pair of isotherms was recorded immediately after the injection, the second pair 1 hour after the end of the recording of the first pair, and so on, for a total of 15 pairs. One pair took 20 minutes to record, the whole series was recorded in 20 hours. The plot indicates that the amount of material on the surface gradually decreases, but remains quite high throughout the series. The initial π at fully open barriers decreases faster than would be possible by subphase water evaporation alone (its typical value is −1 mN/m per day).

In another experiment, the surface layer formed after injection of acetone by the combined method was kept at constant π=25 mN/m and transferred to a polished silicon wafer by horizontal lift. The film can be recognized by the strong blue opalescence after 2 transferred layers. Fig. 9 shows the sample with the 5-layer film erased in part of the area.

Exposure of the sample shown in Fig. 9 to 90-125°C for 55 minutes resulted in significant material loss in the film.

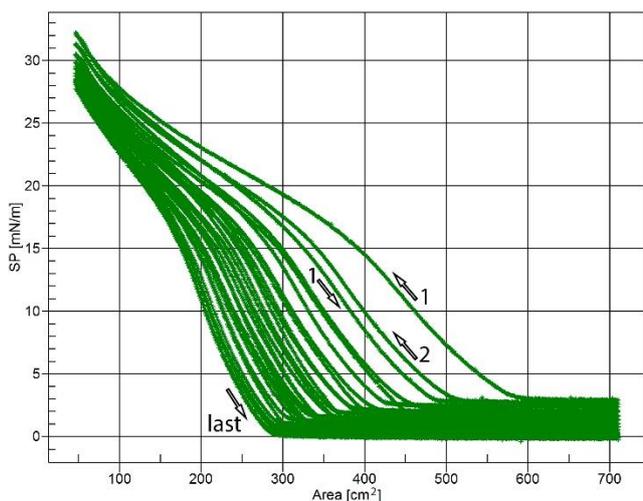

Fig. 8. A series of π-S compression-expansion isotherms of the surface layer of acetone on the aqueous subphase. The first pair of isotherms, the compression branch of the second pair, and the expansion branch of the last pair in the series are marked with arrows.

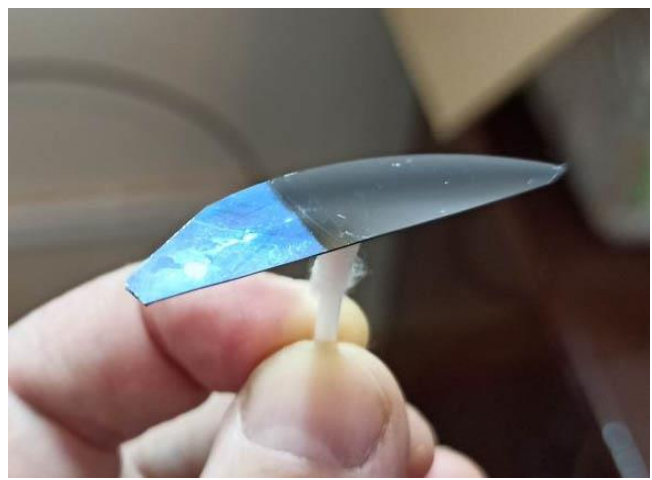

Fig. 9. The acetone film on the polished silicon. 5 layers were transferred from the Langmuir trough at 25 mN/m and the right part of the film was erased. A bright blue opalescence is observed when viewed at a small angle to the plane of the silicon substrate.

### 3.4. Ethanol

The pharmaceutical grade ethanol (95%) was purchased from a local supplier and used as received.

The behavior of ethanol is similar to other solvents studied in this work. A surface layer is formed after injection into the Langmuir trough. The layer is transferable to a substrate.

A fresh surface layer of ethanol is actively losing material, perhaps through evaporation, so the surface pressure may initially decrease even with rapid compression of the layer. To record the π-S isotherm, a substantial amount of ethanol (about 10 mL) should be injected into the Langmuir trough and a long standing time should be allowed for the layer to stabilize. For example, the series of isotherms in Fig. 10 was recorded after dropwise application of 9.8 g of ethanol to a sheet of filter paper 4 cm high and 20 cm wide, with the lower edge in contact with the subphase surface. The first pair of isotherms was recorded 4.5 hours after the ethanol injection, and a 2-hour rest period was allowed before recording each subsequent pair of isotherms.

The barriers of the Langmuir trough were moved at 20 mm/min. This series of isotherms indicates that the amount of material, the hysteresis in the isotherms, and the initial π level at expanded barriers decrease continuously with time.

As a simple visual check of transferability, the surface layer of ethanol held at π=25 mN/m was deposited 70 times on a piece of polished silicon wafer using the horizontal lift method. The part of the wafer was then immersed in water at room temperature. After 18 hours, the film was not dissolved, although its appearance had changed (Fig. 11). With ethanol, it is difficult to produce multi-zone samples similar to those shown in Fig. 4 and Fig. 7 because the initial sharp boundary between zones becomes blurred as additional layers are transferred. Exposing the film to 60-100°C for 20 minutes significantly reduced its contrast with the clean silicon surface.

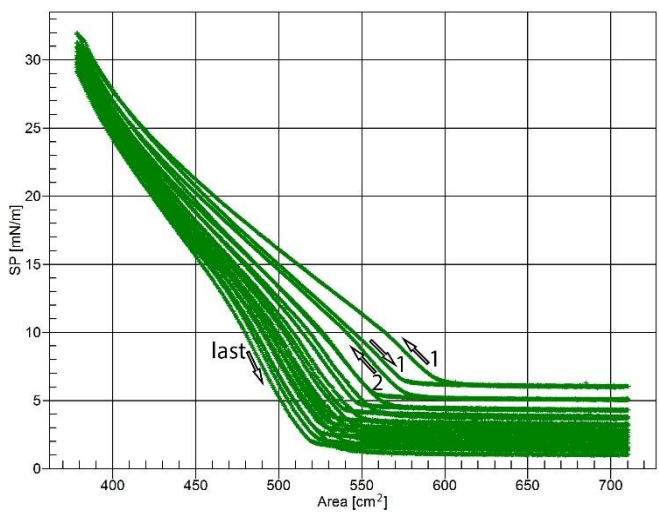

Fig. 10. A series of π-S compression-expansion isotherms of the surface layer of ethanol on the aqueous subphase. The first pair of isotherms, the compression branch of the second pair, and the expansion branch of the last pair in the series are marked with arrows.

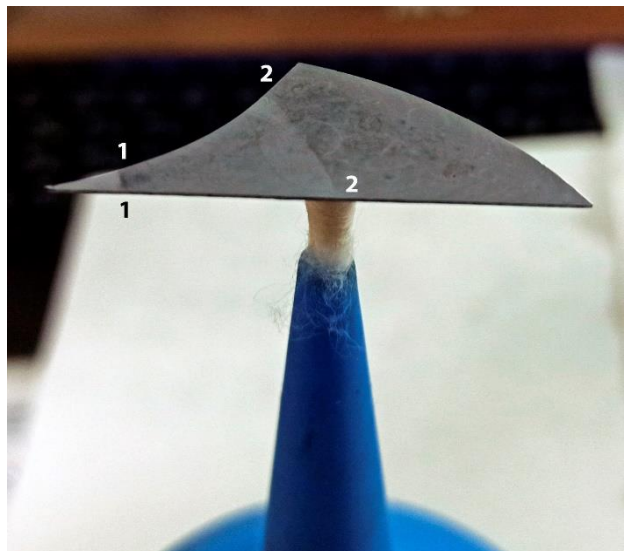

Fig. 11. The film obtained by 70 transfers of the surface layer of ethanol on the silicon substrate at constant π=25 mN/m. The film was erased with a cotton swab left to line 1-1 and immersed in 20 °C water overnight to line 2-2 to test the solubility.

To study the Gibbs layer of ethanol, the solution of 13.3 g ethanol in 774 g water was poured into the Langmuir trough. The surface pressure (π) sensor was zeroed and a series of compression-expansion isotherms were recorded at a barrier speed of 20 mm/min. The first isotherm is almost indistinguishable from zero, but the subsequent isotherms gradually shift to larger S values, and the maximum π achieved in compression increases (Fig. 12). This indicates the gradual increase of the amount of substance in the Gibbs layer of ethanol, similar to DMSO [1]. The initial π at fully expanded barriers decreases faster than would be possible by subphase water evaporation alone.

The Gibbs layer of ethanol can be compressed and repeatedly transferred to a substrate (Fig. 13).

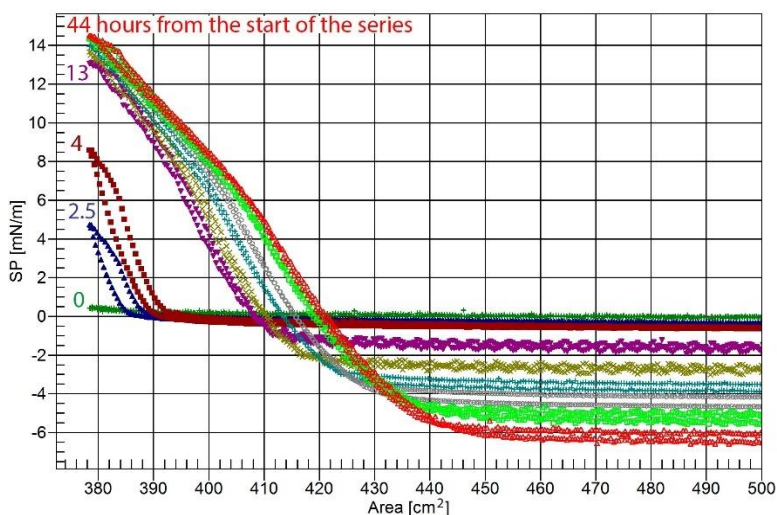

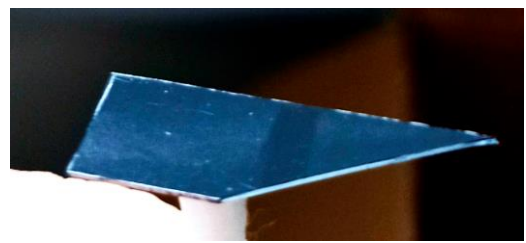

Fig. 12. A series of π-S compression-expansion isotherms of the Gibbs layer on the surface of the aqueous solution of ethanol. The substance gradually accumulates on the surface due to adsorption from the bulk, causing each subsequent isotherm to shift to the right. The time in hours since the start of the isotherm series is shown next to the curve.

Fig. 13. A piece of polished silicon covered with a film obtained by 33 transfers of the Gibbs layer of ethanol at constant π=10 mN/m. The dark stripe near the center of the sample is where the film was erased to compare the remainder of the film to the clean silicon surface.

## 4. Discussion

The experimental data support the validity of our assumption that polar water miscible organic solvents, when injected in the Langmuir trough through the surface of the subphase, form the surface layer which is compressible by barriers and transferable to a substrate. The general picture varies in detail, but is similar for all solvents studied. Immediately after the surface layer is formed, it undergoes significant changes, but over time it stabilizes and then persists for tens of hours with comparatively low losses, even in the case of volatile starting material. The explanation of the surface layer of DMSO (see Introduction section here and [1]) is also applicable to the materials studied in this work. Taking into account volatility and buoyancy, which are absent in DMSO, and the Marangoni flows [11], the significant initial instability and further slow loss of the substance can be understood.

The $\pi$-$S$ isotherm of the surface layer is very similar for different solvents, with two smooth inflections and a region of lower slope in between. The values of $\pi$ corresponding to the inflections also vary little with solvent. This can be interpreted to mean the same contamination in all cases. Of course, all measures have been taken to eliminate any possible source of contamination and to monitor cleanliness, but we still lack categorical evidence that the surface layers originate from the solvents injected into the Langmuir trough. Temporarily accepting such an origin without rigorous proof, one can speculate that the substance-independent nature of the isotherm is caused by a two-dimensional supercritical fluid state of the surface layers of all solvents at room temperature. Such a possibility is shown for ethanol in [12], and other small molecule solvents may behave similarly.

For films on substrates, it is not clear why layer-by-layer transfer is possible (why previous layers do not dissolve when the next layers are transferred) and why the films are so stable despite the volatility of the starting material. Such properties raise the question of whether the films contain the material nominally injected into the Langmuir trough and, if so, in what form the material is present in the film. Various approaches can be used to answer these questions, including chemical analysis, X-ray fluorescence, IR/Raman spectroscopy, isotopic labeling, but each method can be problematic due to the small thickness and very limited amount of material in the film.

The ability of organic solvents to form the transferable surface layer should not be neglected in the preparation and study of LB films. For example, THF is used when working with poorly soluble phthalocyanines [3,5,10,13]. It is quite possible that the observed $\pi$-$S$ isotherm originates from the surface layer of THF, and the very possibility of forming the multilayer film of phtalocyanine may be due to the ability of THF surface layers to be transferred to a substrate.

In [1] we noted that DMSO is the substance with the simplest, smallest, and lightest molecules from which the Langmuir-Blodgett film has ever been prepared. Now ethanol is the winner, with almost no doubt that it will be dethroned by methanol.

## 5. Conclusion

Based on recent results on dimethyl sulfoxide (DMSO) [1], we assumed for general reasons that other polar aprotic water miscible solvents can form surface layers on water that can be compressed by the barriers of the Langmuir trough and transferred to a substrate to form multilayer films. We tested this assumption for four organic solvents, N,N-dimethylacetamide, acetone, ethanol, and tetrahydrofuran, and found it to be true. For each solvent, we worked out the optimal injection mode, studied the surface layers, prepared multilayer films on substrates, and determined their basic properties. This study raised new questions, such as why are the films made from volatile organic compounds stable at room temperature and even when heated.

We plan to verify, refine, supplement, and/or revise the results presented in future versions of this publication.

## 6. Acknowledgements

This work was supported by the Russian Science Foundation (grant 20-12-00175-p), the Ivanovo State University of Chemistry and Technology (studies of the surface layers), and by Ministry of Science and Higher Education of the Russian Federation: projects FZZW-2023-0008 (preparation of substances for investigation) and FZZW-2023-0009 (studies of the multilayer films).